Comment on the paper "Influence of convective boundary condition on double diffusive mixed convection from a permeable vertical surface, by P.M. Patil, E. Momoniat, S. Roy, International Journal of Heat and Mass Transfer, 70 (2014) 313-321"

Asterios Pantokratoras
School of Engineering, Democritus University of Thrace,
67100 Xanthi – Greece
e-mail:apantokr@civil.duth.gr

In the above paper the authors treat the boundary layer flow along a stationary, vertical, permeable, flat plate within a vertical free stream. Fluid is sucked or injected through the vertical plate. The fluid species concentration at the plate is constant and different from that of the ambient fluid. It is also assumed that the plate is heated by convection from another fluid with constant temperature with a constant heat transfer coefficient. The temperature and species concentration difference between the plate and the ambient fluid creates buoyancy forces and the flow is characterized as mixed convection. The partial differential equations of the boundary layer flow ( Eqs. 1-4 in their paper) are transformed and subsequently are solved numerically using an implicit finite difference scheme in combination with a quasi-linearization technique. The quasi-linearization technique is a Newton-Raphson method. The results are presented in 12 figures.

The governing equations and the boundary conditions are as follows

continuity equation:
$$\frac{\partial u}{\partial x} + \frac{\partial v}{\partial y} = 0 \tag{1}$$

momentum equation:
$$u\frac{\partial u}{\partial x} + v\frac{\partial u}{\partial y} = \upsilon\frac{\partial^2 u}{\partial y^2} + g\beta(T - T_\infty) + g\beta^*(C - C_\infty) \tag{2}$$

energy equation:
$$u\frac{\partial T}{\partial x} + v\frac{\partial T}{\partial y} = \frac{\upsilon}{\text{Pr}}\frac{\partial^2 T}{\partial y^2} \tag{3}$$

species equation:
$$u\frac{\partial C}{\partial x} + v\frac{\partial C}{\partial y} = \frac{\upsilon}{Sc}\frac{\partial^2 C}{\partial y^2} \tag{4}$$

$$u = 0, v = v_w, -k\frac{\partial T}{\partial y} = h_f(T_f - T_w), C = C_w \text{ on } y = 0 \tag{5}$$

$$u \rightarrow U_\infty, T \rightarrow T_\infty, C \rightarrow C_\infty \qquad \text{as} \quad y \rightarrow \infty \tag{6}$$

where x and y are the coordinates along and normal to plate, u and v are the velocity components, $\upsilon$ is the fluid kinematic viscosity, β is the fluid thermal expansion coefficient, β* is the species expansion coefficient, T is the fluid temperature, C is the species concentration, g is the gravity acceleration, Pr is the Prandtl number, Sc is the Schmidt number and k is the fluid thermal conductivity. It is assumed that the plate is heated by convection from a fluid with constant temperature $T_f$ with a constant heat transfer coefficient $h_f$. The quantities $U_\infty, T_\infty, C_\infty$ are the velocity, temperature and concentration at the ambient fluid.

The authors transformed the equations (1)-(4) using the following non-dimensional quantities:

$$\xi = \frac{x}{L} \tag{7}$$

$$\eta = \left(\frac{U_\infty}{\upsilon x}\right)^{1/2} y \tag{8}$$

$$G = \frac{T - T_\infty}{T_f - T_\infty} \tag{9}$$

$$H = \frac{C - C_\infty}{C_w - C_\infty} \tag{10}$$

$$F = \frac{u}{U_\infty} \tag{11}$$

$$Gr = \frac{g\beta(T_f - T_\infty)L^3}{\upsilon^2} \tag{12}$$

$$Gr^* = \frac{g\beta^*(C_w - C_\infty)L^3}{\upsilon^2} \tag{13}$$

$$\text{Re}_L = \frac{U_\infty L}{\upsilon} \qquad (14)$$

$$Ri = \frac{Gr}{\text{Re}_L^2} \qquad (15)$$

$$N = \frac{Gr^*}{Gr} \qquad (16)$$

$$A = -2v_w \left( \frac{L}{\upsilon U_\infty} \right)^{1/2} \qquad (17)$$

$$a = \left( \frac{\upsilon}{U_\infty L} \right)^{1/2} \left( \frac{L h_f}{k} \right) \qquad (18)$$

The quantity L is a characteristic length, ξ is the streamwise coordinate, η is the transverse coordinate, G is the temperature, H is the species concentration, F is the velocity, Gr is the Grashof number, Gr* is the species Grashof number, $\text{Re}_L$ is the Reynolds number, Ri is the Richardson number, N is the ratio of species Grashof number to temperature Grashof number, A is the suction-injection parameter and a is a convective parameter. All the above quantities have been taken from the work of Patil et al. (2014) with the same notation.

Something strange was observed in Fig. 8 of Patil et al. (2014) and for that reason the problem was solved again, in the present work, by a different method using the same non-dimensional parameters given by the above equations (7)-(18). Eqs. (1)-(4) represent a two-dimensional parabolic problem. Such a flow has a predominant velocity in the streamwise coordinate which is the direction along the plate. In this type of flow convection always dominates the diffusion in the streamwise direction. Furthermore, no reverse flow is acceptable in the predominant direction. The solution of this problem in the present work is obtained using a finite difference algorithm as described by Patankar (1980). In order to obtain complete form of both the temperature and velocity profiles at the same cross-section, a nonuniform lateral grid has been used. Δy is small values near the surface (dense grid points near the surface) and increases with y. A total of 500 lateral grid cells were used. It is known that the boundary layer thickness changes along x. For that reason, the calculation domain must always be at least equal to or wider than the boundary layer thickness. In each case, the goal was to have a calculation domain wider than the real boundary layer thickness. This has been done by trial and error. If the calculation domain was thin, the velocity and temperature profiles were truncated. In this case a wider calculation domain was used in order to capture the entire velocity and temperature profiles. The parabolic (space marching) solution procedure is described analytically in the textbook of Patankar (1980) which "remains to this day a model of simplicity and clarity and one of the most coherent explications of the finite volume technique ever written" (Acharya and Murthy, 2007). That solution procedure is implicit and unconditionally stable (White, 2006, page 276), has been used extensively in the literature and has been included in fluid mechanics and heat transfer textbooks (see Anderson et al. (1984), p. 364; White (2006), p. 271; and Oosthuizen and Naylor (1999), p. 124). The method has been used successfully in a series of papers by the present author (Pantokratoras, 2009a, Pantokratoras 2009b, Pantokratoras 2010, Pantokratoras, 2014a).

The problem is non-similar and the results depend on the non-dimensional distance $\xi$. The results are presented in Fig. 1 for Pr=0.7, N=0.5, Sc=2.57, Ri=1, a=1, A=1 and different values of distance $\xi$ along the plate. Two temperature profiles which correspond to $\xi=0$ and $\xi=1$ were taken from Fig. 8. of Patil et al. (2014) and three temperature profiles have been produced by the present work for $\xi=0.01$, $\xi=0.1$ and $\xi=1$ for comparison. There are essential differences in the results of the two works. The wall temperature at the plate ($\eta=0$) for $\xi=1$ is 0.875 in Patil et al. (2014) whereas the corresponding value of the present work is 0.6033. The wall temperature for $\xi=0$ is again 0.875 in Patil et al. (2014) whereas the corresponding value of the present work for $\xi=0.01$ is 0.1906. It appears that the wall temperature in the work of Patil et al. (2014) does not depend on $\xi$ and remains constant.

It is well known in convection along a plate with a convective temperature boundary condition that the temperature changes along the plate (see for example Fig. 10 in Pantokratoras, 2014b). This change is caused by the boundary condition given by Eq. (5). However, the wall temperature in the work of Patil et al. (2014) remains constant in contrast to all other works with a convective boundary condition.

Taking into account all the above the credibility of the results presented by Patil et al. (2014) is doubtful.

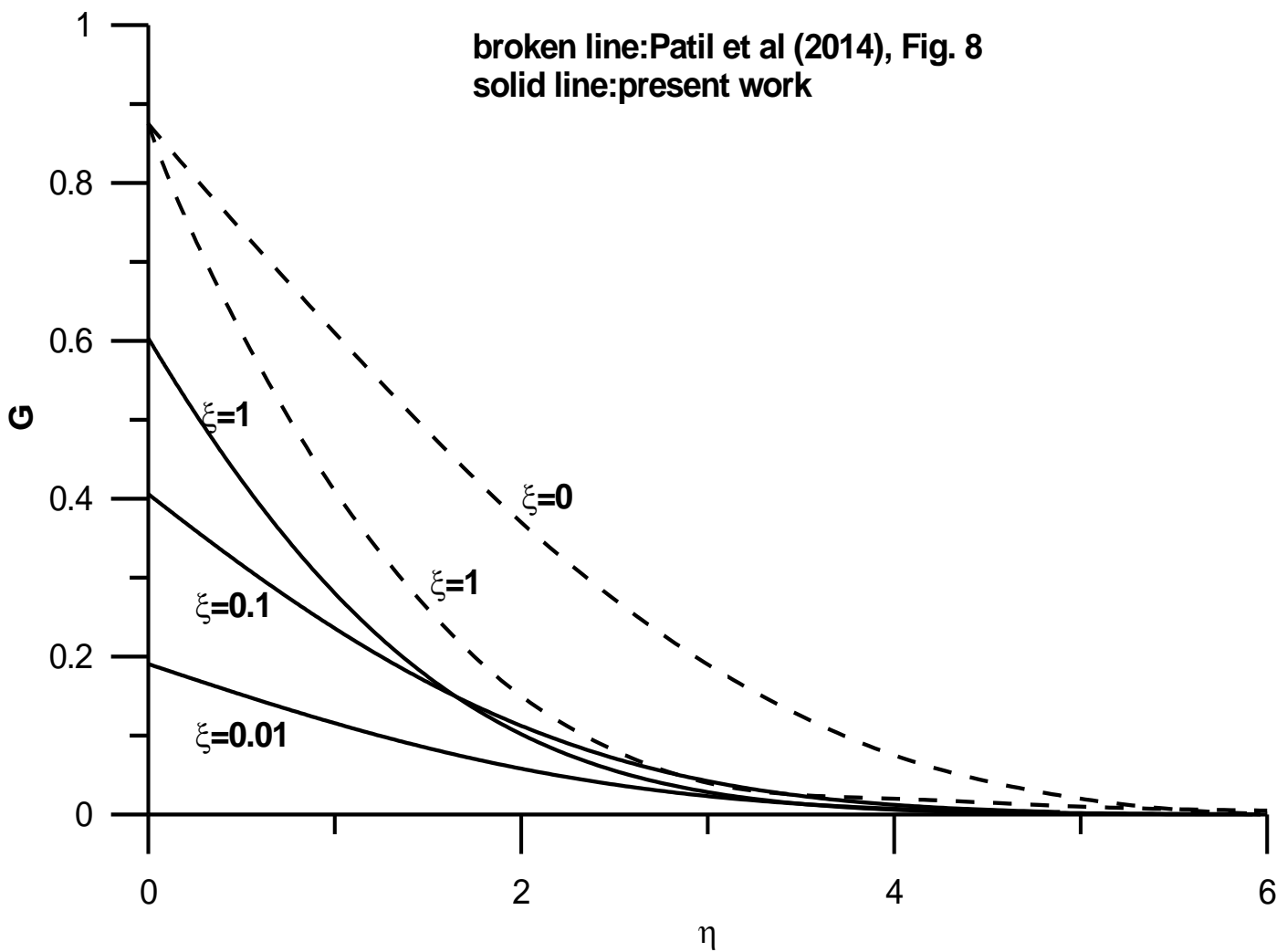

Fig. 1. Effects of $\xi$ on temperature profile when Pr=0.7, N=0.5, Sc=2.57, Ri=1, a=1 and A=1.